\documentclass[prl,aps,amsfonts,amssymb,draft,floats,twocolumn,
showpacs]{revtex4}
\usepackage{epsf}
\newcommand{\be}{\begin{equation}}
\newcommand{\ee}{\end{equation}}
\newcommand{\one}{\openone}

\begin{document}

\title{
Surface Superconductivity of Dirty Two-Band Superconductors:
Applications to ${\rm MgB}_2$.
}

\author{Denis~A.~Gorokhov}

\address{Laboratory of Atomic and Solid State Physics,
Cornell University, Ithaca, NY 14853-2501, USA}

\begin{abstract}
The minimal magnetic field $H_{c2}$ 
destroying superconductivity in the {\it bulk} of a 
superconductor
is smaller than the magnetic field $H_{c3}$ needed to destroy
{\it surface} superconductivity if the surface of the superconductor
coincides with one of the crystallographic planes 
and is parallel to the external magnetic field.
While for a dirty single-band superconductor the ratio of 
$H_{c3}$ to $H_{c2}$
is a universal temperature-independent 
constant 1.6946, for dirty two-band superconductors
this is not the case. I show that in the latter case
the interaction of the two bands leads to a novel scenario
with the ratio $H_{c3}/H_{c2}$ 
varying with temperature and
 taking values larger and smaller than 1.6946.
The results are applied to ${\rm MgB}_2$
and are in agreement with recent experiments [A.~Rydh {\it et al.},
cond-mat/0307445].
\end{abstract} 

\vskip1.5cm

\maketitle

{\it Introduction.} It is well-established that strong magnetic field
destroys superconductivity. If an external field $H$
applied to a type-II superconductor exceeds the second critical field
$H_{c2}$, the {\it bulk} order parameter in the superconductor vanishes.
However, even for $H>H_{c2}$ superconductivity might still exist
in a thin layer close to the surface if $H$ is smaller than the third
critical field $H_{c3}>H_{c2}$\cite{Saintjames}.  In this paper I investigate
the onset of superconductivity via surface nucleation for the field $H$
slightly below the threshold $H_{c3}$.

In their pioneering work\cite{Saintjames} Saint-James and de~Gennes
have shown that if the external magnetic field is applied parallel
to the surface of an isotropic single-band superconductor\cite{footnote_1}
with a temperature close to the transition temperature $T_c$, the ratio 
$H_{c3}/H_{c2}$ takes the universal value $\eta = 1.6946$
independently of the superconducting material.
It turns out\cite{Saintjames} that for $H$ slightly below $H_{c3}$
the superconducting order parameter exists within the distance $\zeta (T)$ 
(the coherence length of the superconductor) from the surface.
For distances exceeding $\zeta(T)$ the order parameter approaches 
zero rapidly. 
The dependence of the ratio $H_{c3}/H_{c2}$ on the material
properties\cite{Saintjames_book,Degennes_book,Strongin},        
sample geometry and topology\cite{Kogan,Saintjames_2,Moshchalkov},  
and temperature\cite{Abrikosov,Kulik,Keller} 
has become a subject of intensive investigations.

A novel window for investigating surface superconductivity was opened
after the discovery of the two-band superconductor 
${\rm MgB}_2$\cite{Nagamatsu}. Not only has it a relatively
high ($\approx 40$~K) $T_c$ but also there exist two different 
superconducting gaps.   
As the consequence of this fact, various
properties of ${\rm MgB}_2$ are quite different from those 
of single-band superconductors. For example, the anisotropy 
$\gamma(T) = H_{c2}^{(ab)}(T)/H_{c2}^{(c)}(T)$ (here, $H_{c2}^{(ab)}(T)$
and $H_{c2}^{(c)}(T)$ stand for the second critical fields in the
$ab$ and $c$-directions respectively; note that the crystal
of ${\rm MgB}_2$ is uniaxial)
of the second critical field ${\rm H}_{c2}$ exhibits strong dependence 
on temperature, see e.g. Ref.~\cite{Angst}. 
For single-band superconductors this ratio is constant.

Another puzzle is that $\gamma (T)$ varies widely 
in different experiments\cite{Welp}. This can be attributed to
the existence of surface superconductivity which might 
affect the observable values of $H_{c2}$ and, hence, the anisotropy.
Consequently, the determination of the third critical field $H_{c3}$
is a very important problem. In a recent experiment\cite{Rydh} it 
has been shown that $H_{c3}/H_{c2}$  for ${\rm MgB}_2$
 might be reduced.

\begin{figure}[t]
\epsfxsize= 0.85\hsize
\epsffile{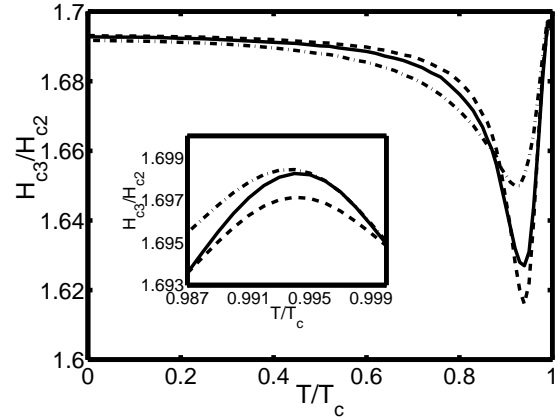}
\caption{\label{third_field}
$H_{c3}/H_{c2}$ 
for the surface of the ${\rm MgB}_2$ crystal coinciding
with the $ab$-plane
and the external magnetic field lying in the $ab$-plane
 for different ratios of the diffusivities 
$D_{2,c}/D_{1,c} = 100, 300, 600$ in the two bands 
(dash-dot, solid, and dashed respectively)
as a function of $T/T_c$.  
Inset: $H_{c3}/H_{c2}$ close to $T/T_c = 1$. 
For $T=T_c$,  $H_{c3}/H_{c2}= 1.6946$.
}
\end{figure}

In the present paper I investigate the ratio $H_{c3}/H_{c2}$ for a
dirty ${\rm MgB}_2$ crystal.
The existence of two different gaps manifests itself through the
remarkable dependence of $H_{c3}/H_{c2}$ on temperature.
This is in sharp contrast with the case of a dirty single-gap superconductor
where $H_{c3}/H_{c2} = \eta$ in the whole temperature range. For
a magnetic field lying in the $ab$-plane 
of the ${\rm MgB}_2$ crystal I find
that if one starts decreasing temperature, $H_{c3}/H_{c2}$ 
first exhibits a maximum at $T\approx 0.99 T_c$ and then a minimum
at $T\approx 0.9 T_c$. As temperature decreases further, $H_{c3}/H_{c2}$ 
increases and tends to a value slightly below $\eta$, 
see Fig.~\ref{third_field}. 
Naively, one could try to use the Ginzburg-Landau 
theory  (GLT)  in order to
find $H_{c3}/H_{c2}$. However, 
as it will be explained below,
this would lead to the ratio
$H_{c3}/H_{c2}= \eta$ in the whole temperature range, i.e.  
one needs a more rigorous approach in order to explain the deviation
of $H_{c3}/H_{c2}$ from $\eta$.   

{\it General formalism.}
An appropriate tool to investigate magnetic properties of
dirty superconductors
is the Usadel equations\cite{Usadel}. For two-band superconductors
they have been derived by Koshelev and Golubov\cite{Koshelev} 
and by Gurevich\cite{Gurevich}. 
Since I investigate
the onset of superconductivity
near $H_{c2}$ or 
$H_{c3}$, it is possible to write the Usadel equations in the linearized
form
\be
\omega f_{\alpha} + \left ( -\sum\limits_{j}\frac{D_{\alpha , j}}{2}
{\left (\nabla_{j} - \frac{2\pi i}{\Phi_0}A_{j}\right )}^2
\right )f_{\alpha} = \Delta_{\alpha}
\label{equation_delta}
\ee
\be
\Delta_{\alpha} = 2\pi T\sum\limits_{\omega > 0}^{\omega_D}
\Lambda_{\alpha\beta}f_{\beta}
\label{equation_f}
\ee
Here, $\omega = 2\pi T (n + 1/2), n = 0,1,\dots$ 
and $\omega_D$ 
are the Matsubara and cutoff phonon frequencies.
$D_{\alpha, j}$ is the diffusion coefficient of the band $\alpha = 1,2$
along the direction $j=x,y,z$. The indices 1 and 2 correspond to the $\sigma$-
and $\pi$-bands respectively. ${\bf A}$ is the vector potential.
$\Delta_{\alpha}$ and $f_{\alpha}$
are the superconducting gap and anomalous green function for
 the band $\alpha$.
The matrix ${\hat \Lambda}$ 
represents the strength of the coupling parameters and
has the values 
$\lambda_{11} \approx 0.81$, $\lambda_{22} = 0.285$, 
$\lambda_{12}\approx 0.119$, and $\lambda_{21}\approx 0.09$,
see Ref.~\cite{Mazin}.
In the present paper I will
concentrate on two geometries: {\it i)} 
the magnetic field is parallel to the $c$-axis and the surface
of the crystal;
{\it ii)} the surface of the superconductor
coincides with the $ab$-plane and the field {\bf H} lies in the $ab$-plane.  
As I will show, in case {\it i} $H_{c3}/H_{c2} = \eta$ at any
temperature. In case {\it ii}, $H_{c3}/H_{c2}$ is shown in
 Fig.~\ref{third_field}. I assume that ${\bf H}$ is parallel to the
$z$-axis.

Choosing the gauge as $A_y = Hx$ I look for the solution
of the form
$f_{\alpha} \equiv f_{\alpha} (\omega , x)\exp\left (
ik_{y}y + ik_{z}z\right )$ and 
$\Delta_{\alpha} \equiv
\Delta_{\alpha} (x) \exp\left (
ik_{y}y + ik_{z}z\right )$.
In general, Eqs.~(\ref{equation_delta}) and (\ref{equation_f})
define a sequence
of solutions corresponding to different eigenvalues $H = H_{c2}$
or $H=H_{c3}$.
One should look for the maximal possible values of 
$H_{c2}$ or
$H_{c3}$. This
corresponds to the case $k_z = 0$. Substituting 
the Ansatz for $f_{\alpha}$ and $\Delta_{\alpha}$ 
into (\ref{equation_delta}) and (\ref{equation_f})
I obtain the system of equations
\begin{widetext}
\be
{\hat \Lambda}^{-1}
\left (
\begin{array}{c}
  \Delta_1 (x) \\
  \Delta_2 (x) 
\end{array} 
\right ) =
\left (
\begin{array}{cc}
2\pi T\sum\limits_{\omega > 0}^{\omega_D}
\frac{1}{\omega + {\hat H}_{1}(x_0)} & 0 \\
0 & 2\pi T \sum\limits_{\omega > 0}^{\omega_D}
\frac{1}{\omega + {\hat H}_{2}(x_0)}
\end{array}
\right )
\left (
\begin{array}{c}
  \Delta_1 (x) \\
  \Delta_2 (x) 
\end{array} 
\right ) ,
\label{main_system}
\ee 
\end{widetext}
where
\be
{\hat H}_{\alpha}(x_0) = 
-\frac{{\tilde D}_{\alpha,x}}{2}\frac{\partial^2}{\partial x^2}
+ \frac{{\tilde D}_{\alpha , z}}{2}
{\left (\frac{2\pi H}{\Phi_0}\right )}^2
{\left (x - x_0 \right )}^2,
\label{operator}
\ee
with ${\tilde D}_{\alpha,x} = {\tilde D}_{\alpha,z} = {D}_{\alpha,a}$
for case {\it i} and
${\tilde D}_{\alpha,x} = D_{\alpha,a}$ and
${\tilde D}_{\alpha,z} = D_{\alpha,c}$ 
for case {\it ii} ($D_{\alpha,a} = D_{\alpha,b}\ne D_{\alpha,c} $
are the diffusion coefficients along the crystallographic axes).
$x_0 = k_y \Phi_0/2\pi H$ is the parameter characterizing how far
away the superconducting nucleus is situated from the surface. 
Note that $x_0$ is the same for the both bands.
 The operator 
(\ref{operator})
can be rewritten in the form 
$\left (\pi H/\Phi_0\right )
\sqrt{{\tilde D}_{\alpha,x}{\tilde D}_{\alpha,z}}
{\hat h}_{\alpha}(x_0^{\prime})$,
with
\be
{\hat h}_{\alpha}(x_{\alpha,0}^{\prime}) = 
- \frac{\partial^2}{\partial {x_{\alpha}^{\prime}}^2} +
{
\left ( 
x_{\alpha}^{\prime} - x_{\alpha,0}^{\prime} 
\right )
}^2,
\ee
where I have made the variable substitution 
$x = \beta_{\alpha} x_{\alpha}^{\prime}$ and 
$x_{0} = \beta_{\alpha} x_{\alpha, 0}^{\prime}$, with 
$\beta_{\alpha} = { \left ( 
{\tilde D}_{\alpha , x}/{\tilde D}_{\alpha , z}\right )}^{1/4}
{ \left (\Phi_0/2\pi H \right )}^{1/2}$.

The system (\ref{main_system})
should be solved with the boundary conditions (BC)
$\partial\Delta_{\alpha}/\partial x|_{x=0} = 0$,
and $\Delta_{\alpha}(x\rightarrow + \infty)\rightarrow 0$,
$\alpha = 1,2$ valid for geometries {\it i} and {\it ii}, see above.
For $H_{c3}$ the BC are well-established for dirty 
superconductors\cite{Degennes_book}. For $H_{c2}$ the application
of these BC gives the same result as the BC requiring the 
appearance of a superconducting nucleus in the bulk.
The procedure of finding
$H_{c3}/H_{c2}$ is following: first, set
$x_0 = 0$ in  (\ref{main_system}) and find the maximal possible
field $H$ for which the solution satisfying the BC exists. This 
gives $H_{c2}$. Next, for $x_0\ne 0$ find the maximal
field $H = H(x_0)$ for which the solution of (\ref{main_system})
exists. Then, $H_{c3} = {\rm max}_{x_0}\left \{H(x_0)\right \}$.
I would like to mention that there are complementary approaches
for calculating $H_{c2}$ based on 
macroscopic theory\cite{Miranovic,Dahm} and GLT\cite{Askerzade,Zhitomirsky}.

Here, it is instructive to study briefly the case of a single-gap
superconductor. This corresponds to $\lambda_{12} = \lambda_{21} = 0$.
$H_{c2}$ and $H_{c3}$ are then determined by those for band 1
(as $\lambda_{11} > \lambda_{22}$). The solution for $\Delta_1 (x)$
is proportional to the ground state wave function of the operator
${\hat H}_1(x_0)$ and 
$\Delta_2 (x) = 0$.
 Substituting this Ansatz into (\ref{main_system}) I obtain the transcendental
equation of the form 
\be
1 - \lambda_{11}\sum_{\omega > 0}^{\omega_D}
\frac{1}{\omega + 
\left (\pi H/\Phi_0\right )
\sqrt{{\tilde D}_{1,x}{\tilde D}_{1,z} }
\epsilon_0(x_{1,0}^{\prime})
} = 0, 
\label{example}
\ee
with $\epsilon_0(x_{\alpha,0}^{\prime})$
the lowest eigenvalue of the operator 
${\hat h}_{\alpha}(x_{\alpha,0}^{\prime})$.
The field $H_{c2}$ can be found as the solution of the above equation
for $x_{1,0}^{\prime} = 0$; note that $\epsilon_0(0) = 1$. 
Assume, a certain value of the magnetic field $H_{c2}$ is found;
let us change the parameter $x_{1,0}^{\prime}$.
This leads to the decrease of the eigenvalue 
$\epsilon_0(x_{1,0}^{\prime})$\cite{Saintjames}. In order to satisfy
Eq.~(\ref{example}) one has to increase the field $H$;
that is why $H_{c3} > H_{c2}$.
The minimal $\epsilon_0(x_{1,0}^{\prime})$  can be realized for 
$x_{1,0}^{\prime} =0.7618$\cite{Saintjames}
 and is equal to  $0.5901$\cite{Saintjames}.
 This means
that $H_{c3}/H_{c2} = 1/0.5901 = \eta$
for any temperature $T$.
Remarkably, it is not necessary
to solve Eq.~(\ref{example}) in order to find the ratio $H_{c3}/H_{c2}$,
although the determination of $H_{c2}$ or $H_{c3}$ alone would require 
the complete analysis.

{\it Case i.}
In this case, 
the ratio $H_{c3}/H_{c2} = \eta$ at all temperatures.
This is a consequence of the fact that the operators ${\hat H}_1(x_0)$
and  ${\hat H}_2(x_0)$  have identical eigenfunctions
(since ${\tilde D}_{1,x}/{\tilde D}_{1,z} 
= {\tilde D}_{2,x}/{\tilde D}_{2,z}=1$).
The functions $\Delta_1(x)$ and $\Delta_2(x)$ are proportional to
the ground state wave function of the operator ${\hat H}_1(x_0)$ 
(or ${\hat H}_2(x_0)$). The equation determining the 
critical fields $H_{c2}$ and $H_{c3}$ has the form
$F\left ( \pi D_{1,a} H\epsilon_0(x_{1,0}^{\prime})/\Phi_0,\thinspace
\pi D_{2,a} H\epsilon_0(x_{2,0}^{\prime})/\Phi_0 \right ) = 0$, with
$F(y_1,y_2)$ a certain function of two arguments. Note that in the
present case $\beta_1 = \beta_2$ and, consequently,  
$x_{1,0}^{\prime} = x_{2,0}^{\prime}$ and 
$\epsilon_{0}(x_{1,0}^{\prime}) = \epsilon_{0}(x_{2,0}^{\prime})$.
The maximal value of $H_{c3}$ can be realized for 
$x_{1,0}^{\prime} = 0.7618$ and is  equal to $\eta H_{c2}$
for all temperatures.

{\it Case ii.}
If the magnetic field lies in the $ab$-plane, the operators
${\hat H}_1(x_0)$ and ${\hat H}_2(x_0)$ 
have different eigenfunctions.
This leads to a complicated transcendental equation depending
on {\it all} eigenvalues of the operators 
${\hat H}_1(x_0)$ and ${\hat H}_2(x_0)$
and not only on the ground state ones. 
Eqs.~(\ref{main_system}) can be solved via expanding functions
$\Delta_1 (x)$ and $\Delta_2 (x)$ over the eigenfunctions of
 the operators
${\hat H}_1(x_0)$ and ${\hat H}_2(x_0)$. 
I have truncated the basis of the operators ${\hat H}_1(x_0)$ and
${\hat H}_2(x_0)$ to subspaces consisting of 70 eigenfunctions
and solved the system (\ref{main_system}) numerically.
In Fig.~\ref{third_field} I show the results of the numerics.
Here, I take
$D_{1,a}/D_{1,c} = 40.0$, $D_{2,a}/D_{2,c} = 0.665$.
These ratios can be obtained using the results for the 
average velocity on the ${\rm MgB}_2$ Fermi surfaces
and assuming isotropic scattering, see Refs.~\cite{Golubov,Brinkman}. 
The ratio $D_{2,c}/D_{1,c}$ takes three values: 100, 300, and 600.
This choice is motivated by the facts that the ratio 
$D_{2,c}/D_{1,c}\approx 100$ can be obtained assuming that
the scattering rate of electrons is the same in both bands.
On the other hand  $R = D_{2,c}/D_{1,c} =  600$ gives a better fit
with experiments on the anisotropy measurements\cite{Golubov}.

The results are shown in Fig.~\ref{third_field}. For temperatures
$T\alt 0.6T_c$
I have found that the ratio $H_{c3}/H_{c2}$ is nearly constant
and has a value slightly below $\eta$. This can be explained 
by the fact that at low temperatures the fields $H_{c2}$ and
$H_{c3}$ are determined mostly by the $\sigma$-band whose coherence
length is much smaller than that of the $\pi$-band. At low $T$
the magnetic field $H_{c2}$ depends on the ground state eigenvalues
of the  operators 
${\hat H}_1(x_0)$ and ${\hat H}_2(x_0)$
and the contribution of excited states is negligible\cite{Gurevich,Koshelev}.
The ratios ${ \left ({\tilde D}_{1,x}/{\tilde D}_{1,z}\right )}^{1/4}$ and
${ \left 
({\tilde D}_{2,x}/{\tilde D}_{2,z}\right )}^{1/4}$  determining the length
$x_0$ are equal to $\approx2.5$ and $0.9$ respectively. 
This means that one can maximize the field $H_{c3}$ by choosing
$x_{1,0}^{\prime} = 0.7618$. The length $x_{2,0}^{\prime}$ 
then is large and the ground-state 
eigenvalue of the operator ${\hat h}_{2}(x_{2,0}^{\prime})$ is 
close to $1$.     
The ratio $H_{c3}/H_{c2}$ then can be calculated
as follows: take the zero temperature expression for 
$H_{c2}$\cite{Gurevich,Golubov} and make there a substitution
$D_{1 , j}\rightarrow D_{1 , j}/\eta$. At low 
temperatures\cite{Gurevich,Golubov},
$
H_{c2} = {\Phi_0 T_c \exp\left (g/2 \right )  }/{2\gamma 
{\left (D_{1,a}D_{1,c}D_{2,a}D_{2,c}\right )}^{1/4} },
$
with
$
g = {\left ({\lambda_0^2}/{w^2} + 
{\ln^2\kappa}/{4}
+{ {2\lambda_{-}}\ln\kappa }/{w}
\right )}^{1/2}
-{\lambda_0}/{w},$ 
$\kappa = {D_{2,a}D_{2,c}}/{D_{1,a}D_{1,c}}$,
$\lambda_{-} = \lambda_{11} - \lambda_{22}$,
$\lambda_0 = {\left (\lambda_{-}^2 + 4\lambda_{12}\lambda_{21}\right )}^{1/2}$,
$w = \lambda_{11}\lambda_{22} - \lambda_{12}\lambda_{21}$,
and $\ln\gamma = 0.5772$.
This procedure yields $H_{c3}/H_{c2} = 1.688, 1.690,$ and $1.691$
for $D_{2,c}/D_{1,c} = 100, 300,$ and $600$ respectively. The values obtained 
in the numerics are slightly larger (but still smaller than $\eta$)
due to a small contribution to the adjustment of $x_0$ from the $\pi$-band. 

If one increases temperature, the ratio $H_{c3}/H_{c2}$
decreases and exhibits a minimum at $T\simeq 0.9T_c$.
Then,  the value of $H_{c3}/H_{c2}$ goes up and takes a maximum
at $T\simeq 0.99 T_c$. At $T=T_c$, $H_{c3}/H_{c2} = \eta$. 
The nontrivial behavior of the ratio $H_{c3}/H_{c2}$ in 
${\rm MgB}_2$ is due to the changing relative importance of the
$\pi$-band. While at low $T$ it is unimportant,
at high $T$ it gives a comparable with the $\sigma$-band 
contribution to the fields $H_{c2}$ and $H_{c3}$.

Let us analyze the situation close to $T_c$ in more detail.
In particular, let us explain why at $T=T_c$, $H_{c3}/H_{c2} = \eta.$ 
For $T_c - T \ll T_c$ the field $H_{c3}$ is small and so are the
eigenvalues of the operators ${\hat H}_1(x_0)$ and ${\hat H}_2(x_0)$,
i.e. one can use the expansion  
$\sum_{\omega > 0}^{\omega_D}
1/{\left (\omega + {\hat H}_{\alpha}(x_0)\right )}
\approx \sum_{\omega > 0}^{\omega_D}1/\omega - \sum_{\omega > 0}^{\omega_D}
{\hat H}_{\alpha}(x_0)/\omega^2
 +\dots$ and
Eqs.~ (\ref{main_system}) can be rewritten in the form
\be
{\hat W}
\left (
\begin{array}{c}
  \Delta_1 (x) \\
  \Delta_2 (x) 
\end{array} 
\right ) = 
\left ( 
\begin{array}{cc}
{\hat R}_1(x_0) & 0 \\
0 & {\hat R}_2(x_0)
\end{array}
\right ) 
\left (
\begin{array}{c}
  \Delta_1 (x) \\
  \Delta_2 (x) 
\end{array} 
\right ), 
\label{new_equations}
\ee
with ${\tilde W} = {\hat \Lambda}^{-1} - 
\ln\left ({2\gamma\omega_D}/{\pi T}\right ) \one_2 $ 
and
${\hat R}_{\alpha}(x_0) = 
2\pi T\sum_{\omega > 0}^{\omega_D}
1/{\left (\omega + {\hat H}_{\alpha}(x_0)\right )}
- 2\pi T\sum_{\omega > 0}^{\omega_D}1/\omega$. $T_c$
is determined by ${\rm det}\thinspace{\hat W} = 0$.
Solving the system (\ref{new_equations}) for 
$\Delta_1(x)$ I obtain
\be
\left (W_{11}{\hat R}_2  (x_0)  + W_{22} {\hat R}_1 (x_0)
- {\hat R}_2   (x_0) {\hat R}_1  (x_0)
\right ) \Delta_1 (x) = 0.
\label{basic_equation}
\ee
Eq.~(\ref{basic_equation}) allows to determine the fields $H_{c2}$
and $H_{c3}$ in a regular way. To lowest order, one can neglect
the term ${\hat R}_2(x_0) {\hat R}_1(x_0)$.

The equation $\left (W_{11}{\hat R}_2(x_0) + W_{22} {\hat R}_1(x_0) \right )
\Delta_1 (x) = 0$ has the ground-state solution 
of the same form as Eq.~(\ref{main_system})  
for a single-gap superconductor (the case $\lambda_{12} = \lambda_{21} = 0$)
with ${\tilde D}_{1,x} \rightarrow D_{X} = W_{22} {\tilde D}_{1,x} + 
 W_{11} {\tilde D}_{2,x}$ and
${\tilde D}_{1,z} \rightarrow D_{Z} = W_{22} {\tilde D}_{1,z} + 
 W_{11} {\tilde D}_{2,z}$ and
the problem 
of finding $H_{c3}$
becomes equivalent to the original one considered by
Saint-James and de Gennes\cite{Saintjames}.
Consequently, to lowest order in $T_c$ the ratio
$H_{c3}/H_{c2}$ has the same value $\eta $ as in the case
 of a single-gap 
superconductor. The approximation described above is equivalent to
the GLT. Consequently, the GLT is unable to explain deviations
of $H_{c3}/H_{c2}$ from $\eta$. 

The maximum of $H_{c3}/H_{c2}$
takes place very close to $T_c$
and is 
at the boundary of the accuracy of the present numerical calculations.
Hence, an analytical approach would be useful.
Temperature corrections to $H_{c3}/H_{c2}$ can be found
by expanding operators ${\hat R}_1(x_0)$ and ${\hat R}_2(x_0)$
to second order in $(T_c - T)$.
One  can decompose the operator in the left-hand side of 
(\ref{basic_equation}) as a sum ${\hat L}_1 (H,x_0) + {\hat L}_2 (H,x_0)$,
with ${\hat L}_1 (H,x_0) 
\propto T_c - T$ and 
 ${\hat L}_2 (H,x_0)\propto {\left (T_c - T \right )}^2$. 
Let 
$|\phi_0\rangle $ be the solution of the equation 
${\hat L}_1 (H,x_0)|\phi_0\rangle = 0$ 
and 
$H = H^{(0)}(x_0)$ the critical field to this order. The correction
to the eigenvalue can be found  
perturbatively and are determined
by the implicit relation
\begin{eqnarray}
\langle \phi_0 | {\hat L}_1 (H,x_0)|\phi_0\rangle + 
\langle \phi_0 | {\hat L}_2 (H^{(0)},x_0)|\phi_0 \rangle = 0.
\label{implicit_relation}
\end{eqnarray}
To this order $x_0 = 0.7618{\left (D_X/D_Z\right)}^{1/4}
{\left (1/2\pi H^{(0)}\right )}^{1/2}$. Temperature corrections to 
$H_{c3}$ due to change in $x_0$ are proportional to 
${\left (T_c - T\right )}^3$ and can be neglected.
Since $|\phi_0\rangle$, ${\hat L}_1 (H,x_0)$, and 
${\hat L}_2 (H,x_0)$ are known, one can find $H_{c3}/H_{c2}$ analytically.
Straightforward but quite cumbersome calculations\cite{Gorokhov}
 show that 
near $T_c$, $H_{c3}/H_{c2}\approx \eta + b (T_c - T)/T_c$,
with $b\simeq 1$ for $D_{2,c}/D_{1,c}$ 
in the range from 100 to 600, in accordance
with the numerics, see inset in Fig.~\ref{third_field}.

{\it Experiment.} Recent experiments\cite{Rydh} show that 
that the ratio $H_{c3}/H_{c2}$ is reduced in the case {\it ii}.
The values $H_{c3}/H_{c2}\approx 1.5$ 
in the temperature range 20--30~K
have been reported.
For the case {\it i} $H_{c3}/H_{c2}\approx 1.7$\cite{Welp,Rydh}. 
The present theory gives that $H_{c3}/H_{c2} = \eta$ for case {\it i}
and $H_{c3}/H_{c2} < \eta$ for case {\it ii}, in agreement with 
\cite{Welp,Rydh}.
Theoretical calculations show\cite{Golubov} that the anisotropy
of ${\rm MgB}_2$ is distributed over a wider temperature range
in an experiment that theory suggests. A somewhat similar situation
takes place in the present work for the ratio $H_{c3}/H_{c2}$. 
There are two main sources of deviation between theory and experiment.
First, surface quality\cite{Hart} might affect $H_{c3}/H_{c2}$.
Second, ${\rm MgB}_2$ is situated somewhere at the boundary of the
applicability of the weak-coupling BCS-theory. It would be
very interesting to repeat the calculation done in the present paper
starting from the Eliashberg equations. 

The method described above can be generalized for an arbitrary
direction of crystallographic axes with respect to the surface
of a superconductor. 
For strongly anisotropic superconductors, surface superconductivity
might disappear if the surface does not
coincide with crystallographic planes\cite{Kogan}. Estimates\cite{Gorokhov} 
show that
${\rm MgB}_2$ is sufficiently anisotropic in order to observe
this kind of effects. The detailed analysis of the onset 
of surface superconductivity in this case is challenging
for both theorists and experimentalists.

In conclusion, I have presented the calculation of the 
ratio $H_{c3}/H_{c2}$ for the two-band superconductor ${\rm MgB}_2$
in the dirty limit. 
Remarkably, in contrast to the
case of a single-gap superconductor, the above ratio is
temperature-dependent. The Ginzburg-Landau theory is unable to
explain deviations of $H_{c3}/H_{c2}$ from $\eta$.

The present work is supported by the Packard Foundation.
D.~A.~G. thanks M.~Angst and A.~E.~Koshelev for 
helpful discussions.


\begin{thebibliography}{99}
\bibitem{footnote_1} In the present paper I assume that
the superconductor fills 
the right half-space ($x>0$) and the left half-space 
($x<0$)
is filled by an
insulator.
\end{thebibliography}
\end{document}